# Evaluation of optimal cut-offs and dichotomous combinations for two biomarkers to improve patient selection

Gina D'Angelo*, Di Ran, Binbing Yu

Oncology Statistical Innovation, AstraZeneca, One Medimmune Way, Gaithersburg, Maryland

*Corresponding author: Gina D'Angelo, AstraZeneca, One MedImmune Way, Gaithersburg, MD 20878; Gina.DAngelo@AstraZeneca.com

# Abstract

**Background:** Identifying the right cut-off for continuous biomarkers in clinical trials is important to identify subgroups of patients who are at greater risk of disease or more likely to benefit from a drug. The literature in this area tends to focus on finding cut-offs for a single biomarker, whereas clinical trials more often focus on multiple biomarkers.

**Methods:** Our first objective was to compare three methods—Youden index, point closest to the (0,1) corner on the receiving operator characteristic curve (ER), and concordance probability—to find the optimal cut-offs for two biomarkers, using empirical and non-empirical approaches. Our second and main objective was to use our proposed logic indicator approach to extend the Youden index and evaluate whether a combination of biomarkers is an improvement over a single biomarker.

**Results:** The logic indicator approach created a condition in which either both biomarkers were positive or only one of the biomarkers was positive. A prostate cancer study and a simulated phase 2 lung cancer study were used to illustrate approaches to finding optimal cut-offs and comparing combined biomarkers with single biomarkers.

**Conclusion:** Our results can aid in determining whether a single biomarker or a combination of biomarkers is superior in identifying patients who are more likely to respond to treatment. This work can be of great importance in the era of personalized medicine, where many treatments do not provide clinical benefit to "average" patients.

**Introduction**

Biomarkers are often used for predictive, prognostic, and screening purposes, as well as to identify patients who are at risk of disease and populations that are likely to benefit from a particular drug. Identifying the right biomarker or set of biomarkers is important for drug development because it can optimize treatment decisions and improve cost-effectiveness. In addition to selecting the right biomarker, it is important to identify a subgroup of patients who are at greater risk of disease or more likely to benefit from a drug, which depends on finding the right cut-off for the biomarker [1].

Finding the optimal cut-off of a biomarker to determine which patients are biomarker positive and negative is of interest in clinical trials and is particularly important in the era of personalized medicine, as many treatments do not provide clinical benefit to the "typical" patient. Moreover, classification of patients is often prone to errors. A healthy patient can be classified incorrectly, or alternatively, one with disease may be declared healthy. The best approach to identify a cut-off is one that optimizes the process and avoids erroneous conclusions.

The approach used to identify the optimal cut-off of a biomarker depends on the criterion of optimality. Many biomarkers, such as PD-L1 and tumor mutation burden, are measured on a continuous scale. Methods used to identify the "optimal" cut-off for continuous biomarkers when dealing with a binary outcome (e.g., disease/no disease, responder/nonresponder) include (1) Youden index, (2) point closest to (0,1) corner on the receiving operating characteristic (ROC) curve (ER), (3) concordance probability, and (4) minimum *P* value. The first three approaches are all based on measurements from the ROC curve and involve functions of sensitivity and specificity. The minimum *P* value approach is used to identify the optimal cut-off by minimizing the *P* value of the Chi-square test on the association between the dichotomized biomarker and true disease status. Rota et al. [2] compared the Youden index, concordance, ER, and minimum *P* value approaches, using a single biomarker demonstrating a normal or gamma distribution. The concordance and ER methods were found to have the best performance. Unal [3] also compared these four methods and added the index of union (IU) approach, also using a single biomarker demonstrating a normal or gamma distribution. The IU approach was found to minimize the summation of absolute values of the differences between the area under the curve (AUC) and specificity and between the AUC and sensitivity. In that study, the IU approach was found to have the best performance. Both Unal [3] and Rota et al. [2] noted that when the data are not normally distributed, the methods may not identify the same cut-off. Thiele et al. [4] evaluated various estimation approaches for the Youden index with data that were normally distributed and skewed, noting that the empirical approach for the Youden index can be noisy and therefore may not always find unique cut-offs.

To date, most methods to determine biomarker cut-offs are limited to a single biomarker [2, 3, 5]. Although many clinical trials consider multiple biomarkers of interest, the performance of these methods has not yet been assessed. We evaluated the performance of the Youden index, ER, and concordance methods in a scenario where there are two biomarkers of interest. These approaches were evaluated under normal and non-normal distributions and with independent and correlated biomarkers. Biomarker cut-offs can be found for different types of binary outcomes such as response/no response and case/control; we used the binary outcome "disease/healthy control."

There is also interest in comparing biomarkers when dichotomized with their cut-offs. Often a single biomarker is insufficient to identify a subpopulation, whereas a combination of biomarkers may provide more accurate diagnosis and stratification. Combinations of biomarkers that include the single biomarker of interest can be correlated, but traditional methods of using both biomarkers in the same model can lead to collinearity issues when modeling approaches are used. We demonstrate a method for comparing two biomarkers as well as a single biomarker with a combination of biomarkers using identified cut-offs for each biomarker. We propose a logic indicator approach whose concept is similar to one developed by Etzioni et al. [6] to determine whether a combination of biomarkers offers benefit over a single biomarker. The Youden index was used as an example to assess the performance of our logic indicator approach.

The focus of our study was to (1) compare methods to identify the optimal cut-offs of two biomarkers and (2) evaluate whether a combination of biomarkers is an improvement over a single biomarker with our proposed logic approach.

**Methods**

**Approaches to determining cut-offs**

ROC curves have been used for decades to demonstrate the classification accuracy of a biomarker. An ROC curve is a graph of the true-positive rate (sensitivity) versus the false-positive rate (1 – specificity) along a range of cut-off levels. The AUC of the ROC is a measure of how well the biomarker can discriminate across all values of the biomarker on average. Because an AUC of 0.5 occurs by chance, the AUC can range from 0.5 to 1. ROC curves are used to select optimal biomarkers and identify their cut-offs. Figure 1 shows an example of an ROC curve. Sensitivity and specificity are determined by a 2×2 table of biomarker (positive/negative) versus disease (yes/no) (Table 1).

Methods used to determine cut-offs are based on the ROC curve, and those used to identify the optimal cut-off that leads to the best disease outcome classification include the Youden index, ER, and concordance probability approaches [2, 3]. Herein sensitivity is denoted by $p(c)$ and specificity is denoted by $q(c)$ at cut-off $c$. The Youden index is the addition of sensitivity plus specificity minus 1, or the difference between the true-positive rate (TPR) and the false-positive rate (FPR) [7]. The Youden index is used to search for the cut-off with the value $c_J$ that maximizes the Youden function $J(c) = q(c) + p(c) - 1$. A biomarker that can perfectly distinguish subjects with disease versus those without disease at cut-off $c$ will have sensitivity $p(c) = 1$ and specificity $q(c) = 1$, with AUC = 1 [8]. A biomarker that is not perfectly able to differentiate disease and non-disease will have $p(c) < 1$ and $q(c) < 1$. The Youden index ranges from 0 when $p(c) = 1 - q(c)$ to 1 when $p(c) = q(c) = 1$. Another approach, point closest to (0,1) on the ROC curve ($ER(c)$), is used to search for the point $c$ that is closest on the ROC curve to (0,1). The point (0,1) represents the ideal situation in which sensitivity and specificity are maximized, i.e., TPR = 1 and FPR = 0, respectively. The point closest to (0,1) is the cut-off that has the smallest Euclidian distance between the ROC curve and the (0,1) point. The equation used for the point closest to (0,1) is

$$ER(c) = \sqrt{(1 - q(c))^2 + (1 - p(c))^2},$$

where the point $c_{ER}$ is found that minimizes *ER(c)*. A third method, concordance probability, is a product of sensitivity and specificity. The equation for concordance probability is $CZ(c) = p(c) * q(c)$. The cut-off maximizing CZ(c) is the optimal cut-off $c_{CZ}$ [5]. The concordance ranges from 0 when $p(c) = 0$ or $q(c) = 0$ to 1 when $p(c) = q(c) = 1$.

These methods may not lead to the same optimal cut-off [2, 5, 8]. Liu [5] and Rota et al. [2] discovered that the same optimal cut-off will be found when the data are normally distributed. Perkins et al. [8] showed that the cut-offs determined by the Youden index and ER differ when $p(c) \neq q(c)$, regardless of the distribution. Operating characteristics are more straightforward to compare when data are normally distributed. However, when data are skewed and have a non-normal distribution, these approaches can lead to different cut-offs, and selection of the best method depends on the question of interest. For example, if a clinical team wants to select a cut-off based on the increase of the true-positive fraction relative to the false-positive fraction, then the Youden index would be selected [9]. The concordance probability would be selected if one wanted to evaluate the probability of being above or below the cut-offs for any pair of diseased and non-diseased subjects [2, 5]. The cut-off approaches described above are referred to as the empirical approach to finding the optimal cut-off. Various alternative approaches, such as using the generalized additive model (GAM) and parametric approaches, can also be used to determine the optimal cut-off.

The Youden index is the only metric approach that uses a normal approximation to find the optimal cut-off. The optimal cut-off is

$$c^* = \frac{(\mu_p \sigma_n^2 - \mu_n \sigma_p^2) - \sigma_n \sigma_p \sqrt{(\mu_n - \mu_p)^2 + (\sigma_n^2 - \sigma_p^2) \log\left(\sigma_n^2 / \sigma_p^2\right)}}{\sigma_n^2 - \sigma_p^2},$$

where the biomarker-negative group (BM−) is BM − $\sim N(\mu_n, \sigma_n^2)$ and the biomarker-positive group (BM+) is BM + $\sim N(\mu_p, \sigma_p^2)$ [4]. The GAM approach is a smoother version of the general linear model, where the function of covariates (here, cut-offs $c_i$) is a smoothed function. Here, GAM is of the form $m_i \sim f(c_i) + \epsilon_i$, where $f$ is a thin plate regression spline, $m$ is the metric value for each cut-off, and $\epsilon_i$ is iid $N(0, \sigma^2)$ [4, 10].

It is also possible to select more than one cut-off when there is no one unique solution. Perkins et al. [8] discussed the Youden index and ER, noting that these methods may not always lead to the same unique cut-off. Based on analytical exploration, this seems to also hold for the concordance probability method. The R package *cutpointr* [11] uses the median as the default to find the cut-off when more than one cut-off is found, as we demonstrate here.

**Combination biomarkers**

Other approaches to combining continuous biomarkers and assessing their performance include comparing AUCs and evaluating linear combinations of the biomarkers; these are common approaches to identifying the best combination of biomarkers [12]. Fewer methods exist when dichotomizing these biomarkers and comparing a single dichotomized biomarker with combined dichotomized biomarkers, and then determining which has the best performance for particular cut-offs. Etzioni et al. [6] used a logic approach with AUC, borrowing the logic regression approach from Ruczinski et al. [13]. We propose a similar logic approach (with ROC-based metrics, e.g. Youden index) that identifies subjects who are positive for either all biomarkers or at least one biomarker and calculated the sensitivity and specificity based on either of these combinations of the "and/∩" or the "or/∪" rule described here.

The "and/∩" rule, using the indicator $I(Y_1 > c_1 \cap Y_2 > c_2)$, specifies that both biomarkers must be positive to calculate sensitivity, and the indicator $I(Y_1 \leq c_1 \cup Y_2 \leq c_2)$ specifies that at least one biomarker must be negative to calculate specificity. This approach has value when a few biomarkers with known cut-offs have been identified as biologically important and there is interest in improving patient selection by determining whether a patient with either or both biomarkers will experience improved efficacy or will be at high risk.

The "or/∪" rule, using the indicator $I(Y_1 > c_1 \cup Y_2 > c_2)$, specifies that either biomarker must be positive to calculate sensitivity, and the indicator $I(Y_1 \leq c_1 \cap Y_2 \leq c_2)$ specifies that both biomarkers must be negative to calculate specificity. The logic approaches use cut-offs for each biomarker to assess whether a combination using "and" or "or" logic is preferred. We demonstrated using the logic indicator with the Youden index and calculate its standard error with the bootstrap. A confidence interval for the Youden index was calculated with the standard error bootstrap to determine the difference between the single biomarkers and the combination of biomarkers. This approach can also be used for the ER and concordance approaches.

Because our objective was to compare biomarkers from the same set of patients, the data were considered to be paired data [14]. Table 2 is arranged by biomarker 1 versus biomarker 2 for the disease and control groups, respectively. The sensitivity and specificity for each biomarker were determined by using the calculation shown in Table 2 [14].

The Youden index, ER, and concordance methods used the sensitivity and specificity of the biomarkers, determined as described above, as well as the formulas shown in Table 2. Cut-offs for each biomarker were used according to the method of choice. Although it is also possible to search for the optimal cut-offs for the combined biomarkers, these may not be similar to those identified for each individual biomarker.

**Case study examples**

Two data sets, one from a prostate cancer study and one from a simulated lung cancer study, were used to identify optimal cut-offs and to compare combined biomarkers with single biomarkers with the methods described above. A training/test data set was applied to all approaches with a 70%/30% split to demonstrate a validation-type approach, although sample size may have been a limiting factor, particularly in the prostate cancer study example. The training data set was used to find the cut-off, and the test data set was used to assess its performance by evaluating sensitivity, specificity, and the cut-off metric (e.g., Youden index and concordance). A bootstrap approach [4, 11] was added to demonstrate a resampling approach and may have improved stability and variability over a training/test data set.

The first data set was from a study [15] in prostate cancer patients and healthy controls to understand early-stage detection. The biomarkers used to assess early-stage disease were free prostate-specific antigen (PSA) (fPSA) and total PSA (tPSA). The second data set was from a lung cancer study that was simulated to mimic a phase 2 trial. This study had responders and nonresponders where the objective response rate was the binary outcome. The objective of the lung cancer study was to identify patients who had a lower rate of disease progression and to determine a cut-off that can be used for late-phase trials. The biomarkers in this study, fluorescence in situ hybridization (FISH) and immunohistochemistry (IHC) results, represent commonly used, related, but distinct continuous biomarkers. Whereas FISH assays the number of gene copies at the DNA level and is often considered to be enriched for true genetic driver events, IHC measures protein expression, which determines biological function. Exploring optimal cut-offs for both biomarkers can potentially improve patient selection. Histograms of the prostate and lung cancer studies (Figures S10 and S11, respectively) show the distribution of biomarkers between the two groups (e.g. disease and healthy control group) in each of these studies and identify where the cut-offs indicate the biomarker-positive and biomarker-negative groups.

**Results**

**Cut-offs for case study examples**

The first example analyzed was a prostate cancer study [15] that had 141 subjects comprising 71 prostate cancer patients and 70 controls. Summary statistics for the biomarkers were as follows. fPSA had a mean ± standard deviation (SD) of 1.8 ± 8.5 and a median (minimum [min], maximum [max]) of 0.58 (0, 100), and tPSA had a mean ± SD of 9.5 ± 19.2 and a median (min, max) of 3.0 (0.03, 100.0). The biomarkers were right-skewed, but transformation did not improve the results because there seemed to be some outliers. The Youden, concordance, and ER approaches all identified the same cut-offs, which were 0.495 for fPSA and 3.3 for tPSA (Table 3). The cut-off values were somewhat higher with the GAM approach but were similar across methods (Table 3). The training/test approach led to (1) smaller cut-offs and worse performance for the empirical Youden index and concordance approaches and (2) smaller cut-offs for the ER-empirical (similar performance) and GAM (improved performance) methods. The cut-offs obtained with bootstrapping were similar to those determined with the empirical approaches.

The second example was a simulated lung cancer study that had 274 subjects comprising 100 responders and 174 nonresponders. Summary statistics for the biomarkers were as follows. FISH had a mean ± SD of 19.5 ± 7.8 and a median (min, max) of 17.3 (9.8, 48.0), and IHC had a mean ± SD of 134.8 ± 28.7 and a median (min, max) of 143.1 (55.0, 169.3). Using the empirical approach, FISH had a cut-off of 14.9 with the Youden index and 17.2 with the concordance and ER approaches (Table 4). The empirical approach had an IHC cut-off of 133.4 with the Youden and concordance approaches and 147.2 with the ER approach, and these values were slightly higher with the GAM approach (Table 4). The training/test approach led to similar results for FISH and IHC; the GAM and Youden approaches yielded larger cut-offs with slightly worse performance. For FISH, the results obtained with bootstrapping were similar to those determined with GAM and were also similar to those for IHC with the GAM and empirical approaches. The data were right-skewed for FISH and left-skewed for IHC. Log transformations for FISH and cubed transformations for IHC were attempted, and the same cut-offs were identified with back-transforming. All approaches identified a unique cut-off for both biomarkers. ER tended to select a higher value for the optimal cut-off. The percentage per subgroup and the sensitivity and specificity can be used to drive a decision in selecting a cut-off.

**Combination biomarkers for case study examples**

We next used our logic indicator approach to determine whether a combination of biomarkers would be an improvement over a single biomarker. For the combined biomarkers, the designation "Comb&" was used when both biomarkers were positive, and the designation "Comb||" was used when only one of the biomarkers was positive. These conditions were calculated for comparison with the single biomarkers, which was done by using the cut-offs obtained with the empirical Youden method. In the prostate cancer study, the cut-offs obtained with the empirical Youden method was used for all approaches because these cut-off methods produced the same values as well as reasonable sensitivity and specificity (0.495 for fPSA and 3.3 for tPSA). We also evaluated the performance of the ER and concordance approaches, using the same cut-offs as those for the Youden index. Table 5 reports the comparison of Youden values between the single/combined biomarkers from the $ith$ column ($column_i$) in Table 5 to each single biomarker (fPSA and tPSA):

$$J_{D,fPSA} = J_{fPSA} - J_{column_i} \text{ and } J_{D,tPSA} = J_{tPSA} - J_{column_i})$$

and its 95% confidence interval with the bootstrapped standard error ($SE_{Boot}$)

$$(J_D \pm z_{1-\frac{\alpha}{2}} * SE_{Boot}, \alpha = .05),$$

where $SE_{Boot} = \frac{1}{B}\sum_{b=1}^{B}\left(\tilde{J}_D^{(b)} - \frac{1}{B}\sum_{r=1}^{B}\tilde{J}_D^{(r)}\right)^2$, $b = 1, \ldots, B$ bootstrap estimates, and $\tilde{J}_D^{(b)}$ is the estimate of the statistic $J_D$ from the $bth$ bootstrap sample. Table 5 also shows the percent positive for each single biomarker and for the combination of biomarkers, along with the sensitivity, specificity, and Youden index for paired data. A statistically significant difference was found between fPSA and tPSA, and tPSA was a better distinguishing biomarker for cancer. Comb& was not a statistically significant biomarker but was preferred over fPSA, whereas tPSA was preferred over Comb& with a statistically significant difference. Comb|| was a statistically significant biomarker and was preferred over fPSA. The ER approach showed a similar trend but not statistically significant difference when comparing both biomarkers to the Comb||. tPSA was preferred over Comb||, but the difference was not statistically significant. We therefore concluded that tPSA was the preferred biomarker and that the combined biomarkers did not result in improved performance, even when both were positive. Another cut-off may show better performance for a combination of biomarkers.

The cut-offs determined by the empirical Youden approach in the lung cancer study (14.9 for FISH and 133.4 for IHC) were used for all approaches because they demonstrated the best sensitivity and specificity. We also evaluated the ER and concordance approaches using the same cut-offs as those for the Youden index. Table 6 reports the comparison of the Youden between the single/combined biomarkers from the $ith$ column ($column_i$) in Table 6 to each single biomarker (FISH and IHC) (i.e., $J_{D,fish}=J_{fish} - J_{column_i}$ and $J_{D,ihc}=J_{ihc} - J_{column_i}$). Also shown is the percent positive for the single and combined biomarkers, as well as the sensitivity, specificity, and Youden index for paired data. No statistically significant difference was found between FISH and IHC, but IHC was preferred over FISH. When each single biomarker was compared with Comb&, the latter was shown to confer a statistically significant improvement over either single biomarker in distinguishing the groups. The Comb|| was not statistically better than each biomarker, but each biomarker was preferred to Comb||. The ER and concordance approaches led to the same conclusions when the cut-off was determined by the Youden index, except the ER method showed a statistically significant difference between FISH and IHC; the lower bound was close to 0, and IHC was preferred. On the basis of these findings, we concluded that Comb& was preferred over a single positive biomarker.

**Discussion**

The determination of biomarkers and their cut-off values has an important impact on which patients receive a diagnosis of disease, which patients are considered to be at risk for disease, and which patients are selected to participate in clinical trials. The literature has focused on identifying cut-offs for a single biomarker, whereas in the clinic, emphasis is more often placed on multiple biomarkers. We focused on the problem of finding a cut-off for two biomarkers. In clinical trials and diagnostic development, there is growing interest in determining whether a combination of dichotomized biomarkers is an improvement over a single biomarker. Here we describe our logic indicator approach to comparing a combination of biomarkers with single biomarkers.

The Youden index, ER, and concordance methods are widely used approaches to identifying biomarker cut-offs. We compared these three methods in the presence of two biomarkers, using a prostate cancer study and a simulated lung cancer study as example data sets. The example data showed that the cut-offs were similar across methods but increased with the non-empirical approaches. In the simulated lung cancer study, the cut-off values did not change much. The use of a higher cut-off value would lead to a smaller pool of candidate patients from which to recruit study participants and would need to be considered for a study or diagnostic tool.

Using the logic indicator approach, we used the example data sets to compare biomarker combinations with single biomarkers. This approach led to a scenario in which either both biomarkers were positive or only one biomarker was positive, being dichotomized with the cut-off found from each biomarker. In the prostate cancer study, tPSA was shown to be a better biomarker for diagnosis than fPSA or a combination of biomarkers. In the simulated lung cancer study, disease progression was better distinguished when both FISH and IHC were positive than when only one biomarker was positive.

In simulation studies, we evaluated normal and non-normal distributions as well as various correlations between the two biomarkers to assess operating characteristics for the Youden, ER, and concordance methods (see Supplemental Material). Both the normally distributed data and the non-normal data showed that the non-empirical methods performed better than the empirical approaches, and that ER-GAM and concordance-GAM performed the best overall in terms of bias and mean squared error.

Classification trees can be used to determine cut-offs for biomarkers and to identify subgroups with multiple cut-offs that predict outcome. In our examples, the cut-offs and the most important independent variable were confirmed with classification trees. Classification trees have limitations, however, including the fact that sensitivity and specificity are not used to identify the cut-offs and that it is not possible to evaluate various types of combinations or to compare a single biomarker with biomarker combinations. Our logic indicator approach is more versatile than classification trees, and these approaches are not comparable. An area for future research will be to extend the use of classification trees to incorporate sensitivity and specificity as well as various types of combinations.

This work focused on finding optimal cut-offs from the marginal distribution even when the biomarkers were correlated. There is also interest in finding the optimal cut-off of multiple biomarkers jointly. This work could be extended to address the classification accuracy of biomarkers jointly, such as by extending the logic indicator approach to find the optimal cut-offs and modeling joint distributions to incorporate the correlation. When multiple cut-offs are identified, they should be evaluated to assess whether one of the cut-offs makes biological sense. A future area for research will be to evaluate other approaches in addition to the median when multiple cut-offs are identified. This work focused on scenarios in which there are two biomarkers, but this approach could be extended and will be addressed in future work. Other future topics to evaluate would be weighting sensitivity and specificity by clinical importance, maximizing sensitivity while fixing sensitivity (and vice versa), extending classification trees, and further development of the logic indicator approach such as a grid search.

**Conclusion**

We compared three methods to find the optimal cut-off in the presence of two biomarkers. We also demonstrated a comparison of a single biomarker to a combination of biomarkers with the cut-off of the biomarker(s) using a logic indicator approach with example data sets. Our results can aid in determining whether a single biomarker or a combination of biomarkers is superior in identifying patients who are more likely to respond to treatment. This work can be of great importance in the era of personalized medicine, where many treatments do not provide clinical benefit to "average" patients.

**Funding statement**

This work was supported by AstraZeneca.

**Conflict of interest statement**

All authors are employees of AstraZeneca and may hold stock ownership, interests, and/or options in the company.

**Author contributions**

GD contributed to the conception and design, method development, identified datasets, analysis and interpretation of data, and drafting of the manuscript. DR contributed to the conception and design, drafting of the manuscript, and interpretation of data. BY contributed to the conception and design. All authors contributed to the review and editing of the manuscript.

**Acknowledgments**

We thank Deborah Shuman of AstraZeneca for editorial support.

**Table 1.** Sensitivity and specificity as determined by a 2x2 table of biomarkers versus disease

| Biomarker | Disease | |
| --- | --- | --- |
| | Yes | No |
| Positive | TP | FP |
| Negative | FN | TN |

TP, true positive; FP, false positive; FN, false negative; TN, true negative.

Sensitivity: TP/(TP + FN); specificity: TN/(TN + FP).

**Table 2.** Biomarker 1 versus biomarker 2 for disease and control groups

| | Disease | | | Controls | | |
|---|---|---|---|---|---|---|
| | Biomarker 1 | | | | Biomarker 1 | |
| Biomarker 2 | Positive | Negative | Biomarker 2 | Negative | Positive |
| Positive | $a_1$ | $c_1$ | Negative | $d_2$ | $b_2$ |
| Negative | $b_1$ | $d_1$ | Positive | $c_2$ | $a_2$ |

Where  $TP_1 = a_1 + b_1$, $FN_1 = c_1 + d_1$, $TN_1 = c_2 + d_2$, $FP_1 = a_2 + b_2$, $SE_1 = TP_1/(TP_1 + FN_1)$, $SP_1 = TN_1/(TN_1 + FP_1)$

$TP_2 = a_1 + c_1$, $FN_2 = b_1 + d_1$, $TN_2 = b_2 + d_2$, $FP_2 = a_2 + d_2$, $SE_2 = TP_2/(TP_2 + FN_2)$, $SP_2 = TN_2/(TN_2 + FP_2)$

**Table 3.** Cut-offs for the prostate cancer study example

| | Youden index | | | | | Concordance | | | | | ER | | | | |
|---|---|---|---|---|---|---|---|---|---|---|---|---|---|---|---|
| Example | Cut-off | % Positive | SE | SP | Metric | Cut-off | % Positive | SE | SP | Metric | Cut-off | % Positive | SE | SP | Metric |
| | Empirical | | | | | | | | | | | | | | |
| fPSA | 0.495 | 0.56 | 0.82 | 0.70 | 0.52 | 0.495 | 0.56 | 0.82 | 0.70 | 0.57 | 0.495 | 0.56 | 0.82 | 0.70 | 0.35 |
| tPSA | 3.3 | 0.48 | 0.80 | 0.84 | 0.65 | 3.3 | 0.48 | 0.80 | 0.84 | 0.68 | 3.3 | 0.48 | 0.80 | 0.84 | 0.25 |
| | Empirical with training/test | | | | | | | | | | | | | | |
| fPSA: training | 0.417 | 0.61 | 0.90 | 0.67 | 0.56 | 0.417 | 0.61 | 0.90 | 0.67 | 0.60 | 0.495 | 0.53 | 0.79 | 0.73 | 0.34 |
| fPSA: test | 0.417 | 0.71 | 0.87 | 0.47 | 0.34 | 0.417 | 0.71 | 0.87 | 0.47 | 0.41 | 0.495 | 0.64 | 0.87 | 0.63 | 0.39 |
| tPSA: training | 2.1 | 0.61 | 0.94 | 0.73 | 0.66 | 2.1 | 0.61 | 0.94 | 0.73 | 0.68 | 3.3 | 0.45 | 0.77 | 0.84 | 0.28 |
| tPSA: test | 2.1 | 0.62 | 0.87 | 0.68 | 0.55 | 2.1 | 0.62 | 0.87 | 0.68 | 0.59 | 3.3 | 0.55 | 0.87 | 0.84 | 0.20 |
| | GAM | | | | | | | | | | | | | | |
| fPSA | 0.664 | 0.44 | 0.68 | 0.79 | 0.46 | 0.673 | 0.44 | 0.66 | 0.79 | 0.52 | 0.673 | 0.44 | 0.66 | 0.79 | 0.40 |
| tPSA | 4.6 | 0.40 | 0.66 | 0.87 | 0.53 | 4.4 | 0.40 | 0.68 | 0.87 | 0.60 | 4.4 | 0.40 | 0.68 | 0.87 | 0.35 |
| | GAM with training/test | | | | | | | | | | | | | | |
| fPSA: training | 0.525 | 0.52 | 0.77 | 0.73 | 0.50 | 0.567 | 0.48 | 0.71 | 0.73 | 0.51 | 0.573 | 0.47 | 0.69 | 0.73 | 0.42 |
| fPSA: test | 0.525 | 0.64 | 0.87 | 0.63 | 0.50 | 0.567 | 0.60 | 0.87 | 0.74 | 0.64 | 0.573 | 0.60 | 0.87 | 0.74 | 0.29 |
| tPSA: training | 3.7 | 0.40 | 0.73 | 0.86 | 0.59 | 3.7 | 0.40 | 0.73 | 0.86 | 0.63 | 3.7 | 0.40 | 0.73 | 0.86 | 0.30 |
| tPSA: test | 3.7 | 0.52 | 0.83 | 0.84 | 0.67 | 3.7 | 0.52 | 0.83 | 0.84 | 0.70 | 3.7 | 0.52 | 0.83 | 0.84 | 0.23 |
| | Bootstrap | | | | | | | | | | | | | | |
| fPSA | 0.476 | 0.57 | 0.82 | 0.69 | 0.50 | 0.513 | 0.55 | 0.80 | 0.70 | 0.56 | 0.533 | 0.54 | 0.79 | 0.71 | 0.36 |
| tPSA | 3.0 | 0.50 | 0.80 | 0.80 | 0.60 | 3.1 | 0.50 | 0.80 | 0.82 | 0.67 | 3.1 | 0.50 | 0.80 | 0.83 | 0.26 |

SE, [sensitivity]; SP, [specificity].

**Table 4.** Cut-offs for the simulated lung cancer study example

| Example | Youden index | | | | | Concordance | | | | | ER | | | | |
|---|---|---|---|---|---|---|---|---|---|---|---|---|---|---|---|
| | Cut-off | % Positive | SE | SP | Metric | Cut-off | % Positive | SE | SP | Metric | Cut-off | % Positive | SE | SP | Metric |
| | Empirical | | | | | | | | | | | | | | |
| FISH | 14.9 | 0.72 | 0.88 | 0.37 | 0.25 | 17.2 | 0.51 | 0.66 | 0.57 | 0.37 | 17.2 | 0.51 | 0.66 | 0.57 | 0.55 |
| IHC | 133.4 | 0.57 | 0.80 | 0.55 | 0.35 | 133.4 | 0.57 | 0.80 | 0.55 | 0.44 | 147.2 | 0.46 | 0.66 | 0.65 | 0.49 |
| | Empirical with training/test | | | | | | | | | | | | | | |
| FISH: training | 15.4 | 0.71 | 0.87 | 0.38 | 0.25 | 17.5 | 0.50 | 0.62 | 0.57 | 0.35 | 17.9 | 0.48 | 0.62 | 0.59 | 0.56 |
| FISH: test | 15.4 | 0.61 | 0.75 | 0.48 | 0.23 | 17.5 | 0.45 | 0.61 | 0.65 | 0.39 | 17.9 | 0.43 | 0.57 | 0.65 | 0.56 |
| IHC: training | 133.4 | 0.59 | 0.81 | 0.52 | 0.33 | 146.8 | 0.50 | 0.72 | 0.61 | 0.44 | 146.8 | 0.50 | 0.72 | 0.61 | 0.48 |
| IHC: test | 133.4 | 0.53 | 0.79 | 0.60 | 0.39 | 146.8 | 0.41 | 0.57 | 0.68 | 0.39 | 146.8 | 0.41 | 0.57 | 0.68 | 0.54 |
| | GAM | | | | | | | | | | | | | | |
| FISH | 16.0 | 0.62 | 0.75 | 0.45 | 0.19 | 17.3 | 0.50 | 0.64 | 0.57 | 0.36 | 17.5 | 0.48 | 0.63 | 0.59 | 0.55 |
| IHC | 134.4 | 0.57 | 0.78 | 0.55 | 0.33 | 146.3 | 0.48 | 0.67 | 0.62 | 0.42 | 146.8 | 0.47 | 0.67 | 0.63 | 0.49 |
| | GAM with training/test | | | | | | | | | | | | | | |
| FISH: training | 16.0 | 0.65 | 0.78 | 0.41 | 0.19 | 17.3 | 0.52 | 0.66 | 0.54 | 0.36 | 17.5 | 0.50 | 0.63 | 0.57 | 0.57 |
| FISH: test | 16.0 | 0.55 | 0.68 | 0.52 | 0.20 | 17.3 | 0.46 | 0.61 | 0.63 | 0.38 | 17.5 | 0.45 | 0.61 | 0.65 | 0.53 |
| IHC: training | 140.8 | 0.54 | 0.73 | 0.56 | 0.30 | 143.2 | 0.52 | 0.73 | 0.59 | 0.43 | 145.8 | 0.52 | 0.73 | 0.60 | 0.48 |
| IHC: test | 140.8 | 0.48 | 0.68 | 0.62 | 0.30 | 143.2 | 0.44 | 0.61 | 0.64 | 0.39 | 145.8 | 0.42 | 0.57 | 0.66 | 0.55 |
| | Bootstrap | | | | | | | | | | | | | | |
| FISH | 15.7 | 0.65 | 0.78 | 0.42 | 0.20 | 17.9 | 0.47 | 0.60 | 0.61 | 0.37 | 17.6 | 0.47 | 0.60 | 0.60 | 0.56 |
| IHC | 137.2 | 0.56 | 0.76 | 0.55 | 0.31 | 138.9 | 0.55 | 0.74 | 0.54 | 0.40 | 141.7 | 0.51 | 0.71 | 0.59 | 0.50 |

SE, [sensitivity]; SP, [specificity].

**Table 5.** Youden differences between biomarkers and combined biomarkers for the prostate cancer example

| Example | Youden difference (95% CI) | | |
| --- | --- | --- | --- |
| | tPSA | Comb& | Comb\|\| |
| fPSA | **−0.129 (−0.254, −0.003)** | −0.058 (−0.163, 0.046) | **−0.070 (−0.130, −0.011)** |
| tPSA | — | **0.070 (0.012, 0.129)** | 0.058 (−0.047, 0.163) |
| Example | % Positive | SE | SP | Youden index |
| --- | --- | --- | --- | --- |
| fPSA | 0.56 | 0.82 | 0.70 | 0.517 |
| tPSA | 0.48 | 0.80 | 0.84 | 0.646 |
| Comb& | 0.45 | 0.73 | 0.84 | 0.575 |
| Comb\|\| | 0.59 | 0.89 | 0.70 | 0.587 |

CI, confidence interval; SE, [sensitivity]; SP, [specificity].

**Table 6.** Youden differences between biomarkers and combined biomarkers for the simulated lung cancer study

|  | Youden difference (95% CI) | | |
| --- | --- | --- | --- |
| **Biomarker** | **IHC** | **Comb&** | **Comb\|\|** |
| FISH | −0.052 (−0.201, 0.097) | **−0.181 (−0.289, −0.074)** | **0.129 (0.049, 0.210)** |
| IHC | — | **−0.129 (−0.210, −0.049)** | **0.181 (0.073, 0.290)** |

| **Biomarker** | **% Positive** | **SE** | **SP** | **Youden index** |
| --- | --- | --- | --- | --- |
| FISH | 0.72 | 0.89 | 0.38 | 0.263 |
| IHC | 0.57 | 0.78 | 0.53 | 0.315 |
| Comb& | 0.43 | 0.72 | 0.73 | 0.444 |
| Comb\|\| | 0.87 | 0.95 | 0.18 | 0.134 |

CI, confidence interval; SE, [sensitivity]; SP, [specificity].

**Figure 1.** ROC curve demonstrating Youden, ER, and concordance.

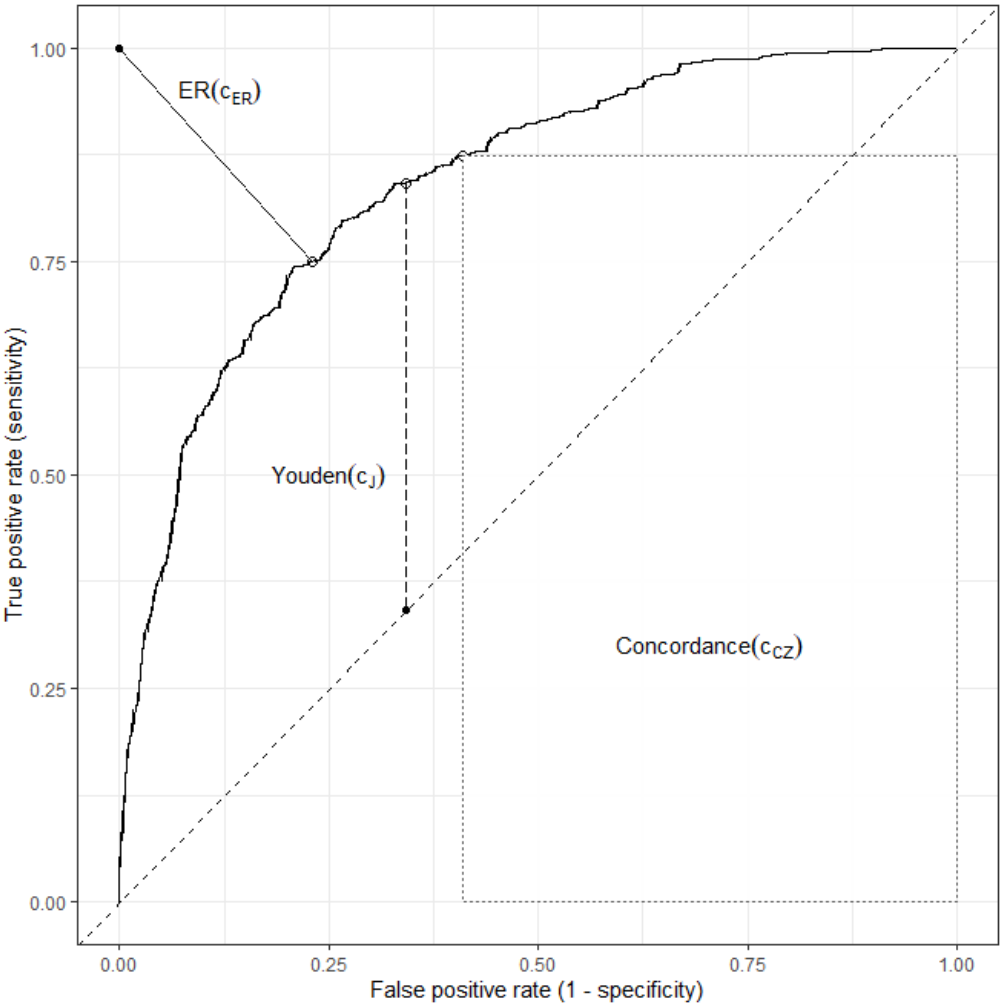

# Evaluation of the optimal cut-offs and dichotomous combinations for two biomarkers to improve patient selection

Gina D'Angelo, Di Ran, Binbing Yu

**Supplemental Material**

## Simulation studies

We performed simulation studies to compare cut-off metrics for two biomarkers when the distribution is normally distributed. The approaches assessed were:

1. Empirical approaches:
    Youden
    Concordance
    Point closest to the (0,1) corner on the receiving operator characteristic (ROC) curve (ER method)

2. Non-empirical approaches:
    Youden-normal
    Concordance–generalized additive model (GAM)
    ER-GAM

Using non-normally distributed data, the same methods were compared, except the Youden-normal was replaced with the Youden-GAM. We evaluated a number of scenarios with simulation studies to assess the statistical properties of the methods. Each simulation study had 10,000 replications. We assessed the following summary statistics: relative bias, mean squared error (MSE), and 95% coverages. The bias is the difference between the estimated cut-off value from each replicate and the true cut-off, where $bias = (\hat{c}_s - c)$, $c$ is the true cut-off, $\hat{c}_s$ are the estimated cut-offs from each replicate s=s,…S, and S is the total number of replicates from the simulation study. The relative bias (RB) is the average of the bias divided by the true cut-off multiplied by 100, where $RB = 100 * \frac{1}{S}\sum_{s=1}^{S}\frac{(\hat{c}_s - c)}{c}$. The MSE is $MSE = \frac{1}{S}\sum_{s=1}^{S}(\hat{c}_s - c)^2$. The 95% coverage is the percentage of replications that are between the 0.025 and 0.975 normal variates. Different distributions, varying sample sizes with different ratios across groups, and varying sensitivity and specificity were evaluated. Both normal and skewed-normal distributions were evaluated. Samples sizes were 50, 100, and 200 per group with a 1:1 ratio and a 1:2 ratio for disease to control. Normal distribution results are shown here. The terms "marker" and "biomarker" are used interchangeably throughout.

To generate bivariate normal variates: controls are $(X_c, Y_c) \sim BVN(\mathbf{0}, \Sigma_c)$ where $\Sigma_c = \begin{pmatrix} 1 & \rho \\ \rho & 1 \end{pmatrix}$, and diseased are $(X_d, Y_d) \sim BVN(\boldsymbol{\mu_d}, \Sigma_d)$ where $\boldsymbol{\mu_d}$ have different values for each scenario, and $\Sigma_d = \begin{pmatrix} 1 & \rho \\ \rho & 1 \end{pmatrix}$. Correlations considered are $\rho = \{0, 0.25, 0.5, 0.75\}$. For the normal distribution of the diseased group, the following means were used: $\mu = (0.51, 1.05, 1.68, 2.56)$ and cut-offs (0.25, 0.52, 0.84, 1.28), respectively. When the variables are independent, i.e., $\rho = 0$, two $N(\mu, 1)$ variables are generated where the μ are specified above for the control and diseased groups. Marker 1 was kept at a $\mu = 0.51$ for each simulation study to assess what happens when the second biomarker has a larger mean for the disease than the controls, e.g., simulation 1 has marker 2 with $\mu = 1.05$ and simulation 2 has Marker 2 with $\mu = 1.68$, etc.

Based on these mean and cut-off values, the sensitivity and specificity are similar and approximately (0.6, 0.7, 0.8, 0.9). The theoretical optimal cut-off occurs where the pdfs of the control and disease groups intersect. Therefore, the optimal cut-off is found by calculating the sensitivity and specificity by $SE = P(X > c)$ using the disease group and $SP = P(X < c)$ using the control group, where $c$ is the cut-off value. Plots (Figure S9) are provided to demonstrate the biomarker distribution of the two groups (e.g., disease and healthy control group) and where the cut-off would indicate biomarker-positive and -negative groups.

The normally distributed biomarker results from the simulation study of 10,000 replicates are shown in Figures S1–S4. Results are shown for the normally distributed data with no correlation, $\rho = 0$, that compare Marker 1 with $\mu = 0.51$ to Marker 2 with any specified mean (the results shown are from Marker 2 with $\mu = 1.05$, but the results are the same for any specified mean for Marker 2), Marker 2 with $\mu = 1.05$ compared to Marker 1 with $\mu = 0.51$, Marker 2 with $\mu = 1.68$ compared to Marker 1 with $\mu = 0.51$, and Marker 2 with $\mu = 2.56$ compared to Marker 1 with $\mu = 0.51$. Figure S1 shows the RB when both markers are normally distributed. The RB decreases as sample size increases and is largest when the sample size is imbalanced. The RB decreases as there is more separation between disease and controls for the second biomarker, i.e., as the mean increases for the second biomarker. The Youden-normal approach has the smallest RB, and Concordance-GAM and ER-GAM have larger RBs but the next-smallest RB. ER and Concordance tend to have the largest RB. An exception occurs with Youden having the largest RB when the biomarker has a small mean and does not have much separation between disease and control such as with Marker 1; however, when there is sample size imbalance, Youden has less RB than Concordance and ER.

Figure S2 shows the MSE when both markers are normally distributed. The MSE decreases as sample size increases; however, the MSE is smaller with the imbalanced sample size (50/100) than a sample size of 50/50 but has a larger MSE than 100/100 and 200/200. Youden has the largest MSE consistently across all markers and sample sizes. Concordance tends to have the next-largest MSE except for Marker 1, where the mean is the smallest. Youden-normal has the second-highest MSE with Marker 1 but the smallest MSE with Marker 2, meaning that the Youden-normal has the best MSE performance as the disease and controls have more separation. The ER-GAM, Concordance-GAM, and ER have the next-smallest MSE for the second markers where the mean increases, and a more obvious separation of the methods emerges as the mean increases showing the non-empirical approaches with better performance.

Figure S3 shows the coverages when both markers are normally distributed. The coverages are around 95% for all methods across all sample sizes, except for Youden-normal with Marker 1 when the sample sizes are smaller than 200. This means that when the disease and control have less separation and the sample size is less than 200/group, the coverages for the Youden-normal will be too large and conservative. It appears that when there is minimal bias and larger variance, the coverages tend to be more conservative. When Marker 2 approaches a large mean with a great deal of separation between groups and there is sample size imbalance, the coverage decreased a small amount for Concordance (although this is a minimal change as it is still close to 95% coverage).

Figure S4 shows the frequency of cut-offs across methods. The Youden method has the most number of cut-offs and the Youden-normal consistently has 1 cut-off. The normal estimation is the reason for a single cut-off being identified. As a result of its function, the Youden can find multiple optimal cut-offs. As the sample size increases, more cut-offs are identified, particularly with the ER-GAM and Concordance-GAM approaches. As the mean of the markers increases, the number of cut-offs reduces for the ER-GAM and Concordance-GAM approaches but slightly increases for the concordance and ER approaches. It appears that the smoothed function of the GAM increases the chances of finding more than 1 cut-off when there are more data.

The normally distributed markers showed the non-empirical methods to have the best performance overall. Youden-normal has the best performance when there is more separation between the healthy and diseased group (all second markers), and the ER-GAM and Concordance-GAM had the next-best performance overall. Regarding the empirical approaches, Youden has the smallest RB but larger MSE caused by more variation, whereas the ER has the largest RB but smallest MSE. The skewed normally distributed markers showed the ER-GAM and Concordance-GAM methods to have the best performance overall. The correlations did not have much impact on the results for the empirical and non-empirical approaches for both distributions. This makes sense because the correlations would not shift the marginal distributions and therefore would not change the cut-offs.

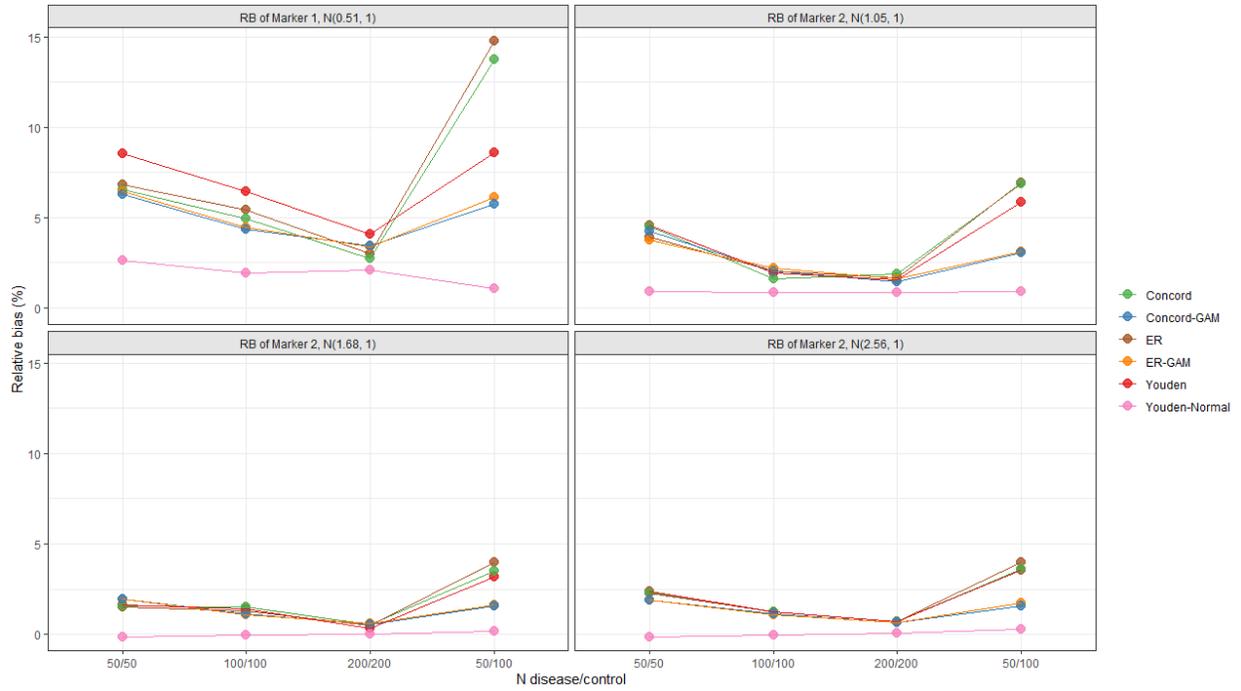

**Figure S1.** Relative bias of normally distributed markers with no correlation between Marker 1 and 2.

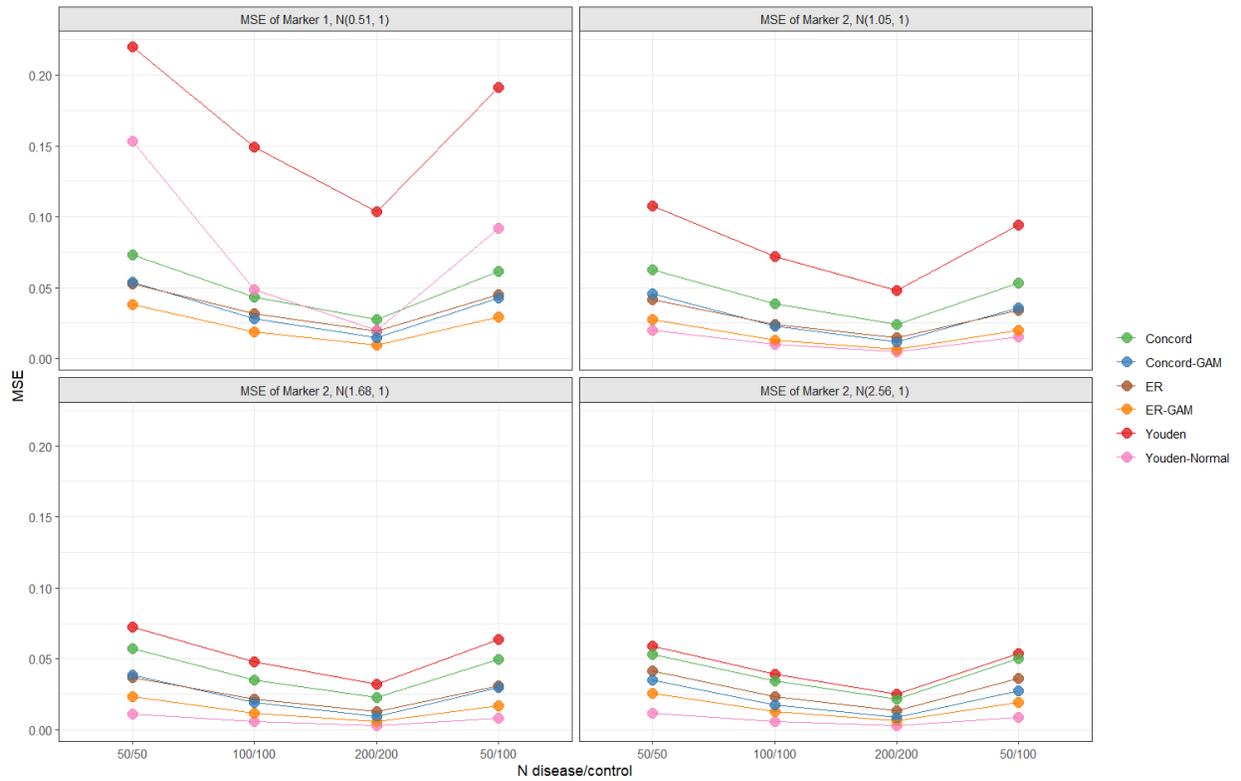

**Figure S2.** MSE of normally distributed markers with no correlation between Marker 1 and 2.

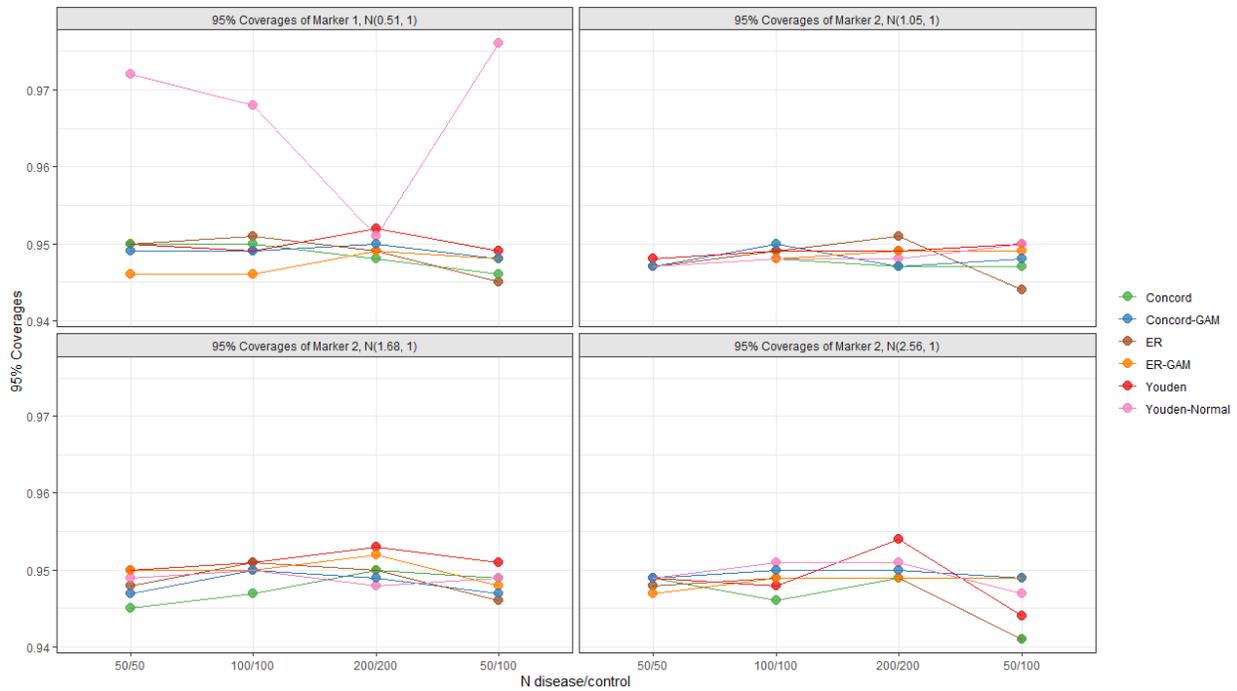

**Figure S3.** 95% Coverages of normally distributed markers with no correlation between Marker 1 and 2.

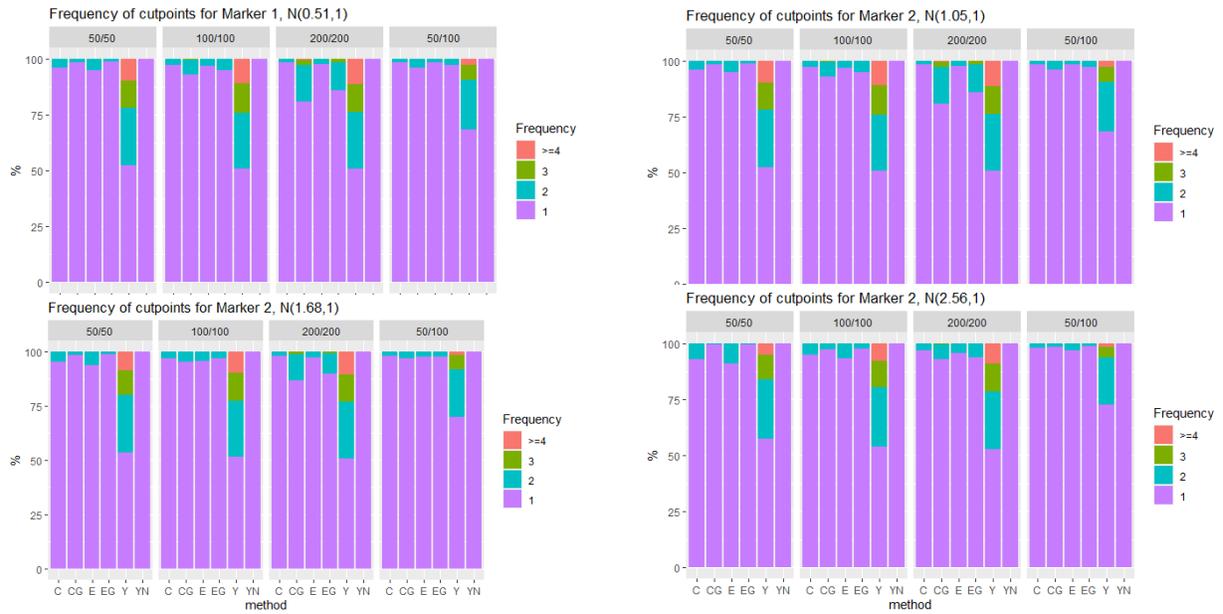

**Figure S4.** Frequency of cut-offs when Marker 1 and 2 are normally distributed with no correlation. Methods: C, Concordance; CG, Concordance-GAM; E, ER; EG, ER-GAM; Y, Youden; YN, Youden-normal.

## Simulation results for skewed biomarkers

We performed simulation studies to compare these cut-off metrics for two biomarkers: Youden, Concordance, ER, Youden-GAM, Concordance-GAM, and ER-GAM when the distribution is a skewed normal distribution. We evaluated a number of scenarios with simulation studies to assess the statistical properties of the methods. Each simulation study has 10,000 replications. We assessed the following summary statistics: average of the bias, RB, MSE, and 95% coverages. Different distributions, varying sample sizes with different ratios across groups, and varying sensitivity and specificity were evaluated. In this supplement the skewed normal distribution is evaluated. Sample sizes considered are 50, 100, 200 per group with a 1:1 ratio and 1:2 ratio for disease to control.

A skewed normal distribution $SN(\xi, \omega, \alpha)$ was used to create a skewed distribution where $\xi$ is the location parameter, $\omega$ is the scale parameter, and $\alpha$ is the slant/shape parameter. A mean, variance, and skewness are assigned and converted to the SN parameters where the mean is

$$\mu = \xi + \omega\delta\sqrt{\frac{2}{\pi}}, \delta = \frac{\alpha}{\sqrt{1+\alpha^2}}, \text{variance is } \sigma^2 = \omega^2\left(1 - \frac{2\delta^2}{\pi}\right),$$

and skewness is

$$\gamma_1 = \frac{4-\pi}{2} \frac{\left(\delta\sqrt{2/\pi}\right)^3}{(1-2\delta^2/\pi)^{3/2}}.$$

The cut-offs for the skewed normal distribution vary across the methods and are identified by finding the cut-off value that maximizes the Youden and concordance and minimizes the ER. The theoretical optimal cut-off occurs where the pdf of the control and disease groups intersect. Therefore, the optimal cut-off is found by calculating the sensitivity and specificity by: $SE = P(X > c)$ using the disease group and $SP = P(X < c)$ using the control group where $c$ is the cut-off value.

Two bivariate skewed correlated normal variates $(X_g, Y_g) \sim BSN(\xi_g, \Omega_g, \alpha_g)$ were also evaluated for the control and the disease groups where BSN indicates bivariate skewed normal, $g = c$ indicates controls and $g = d$ indicates diseased group. The controls are fixed across all scenarios while we evaluate different means for the disease group to assess the metrics with varying mean and skewness values. A mean, variance, and skewness $(\mu, \Sigma, \gamma)$, where $\Sigma_g = \begin{pmatrix} 1 & \rho \\ \rho & 1 \end{pmatrix}$, are assigned to the disease and control group and converted to the BSN parameters $(\xi_g, \Omega_g, \alpha_g)$ using the *cp2dp* function from the R package *SN* [1]. The mean and skewness for the controls are $\mu = (0,0)$ and $\gamma = (0.3, 0.4)$, and for the diseased group are $\mu =$

{ (0.55, 1), (0.55, 2), (0.55, 3) } and $\gamma = (0.4, 0.5)$. Other scenarios for the disease group mean are $\mu = \{ (1, 2), (1, 3)\}$ and $\gamma = (0.4, 0.5)$. The various mean values consider varying sensitivity and specificity from 0.5–0.9. Table S1 shows the cut-offs for each method for the various scenarios. Correlations are $\rho = \{0, 0.25, 0.5, 0.75\}$. When the variables are independent, i.e., $\rho = 0$, two skewed normal variables are generated where the mean, variance, and skewness are specified above for the control and disease group. Marker 1 was kept at $(\mu, \gamma) = (0.55, 0.4)$ for the disease group at each simulation study to assess what happens as the second biomarker has a larger mean for the disease group compared to the controls, e.g., simulation 1 has Marker 2 with $(\mu, \gamma) = (1, 0.5)$, simulation 2 has Marker 2 with $(\mu, \gamma) = (2, 0.5)$, etc.

**Table S1.** Cut-offs for a skewed normal distribution

| Control, disease $\mu, \gamma$ | $c_J$, (SE, SP) | $c_{cz}$, (SE, SP) | $c_{ER}$, (SE, SP) | $J(c_J)$ | $C(c_{cz})$ | $ER(c_{ER})$ |
|---|---|---|---|---|---|---|
| $\mu = (0, 0.55), \gamma = (0.3, 0.4)$ | 0.07, (0.67, 0.55) | 0.17, (0.63, 0.59) | 0.19, (0.62, 0.60) | 0.33 | 0.37 | 0.56 |
| $\mu = (0, 1), \gamma = (0.3, 0.4)$ | 0.32, (0.74, 0.64) | 0.39, (0.72, 0.67) | 0.41, (0.71, 0.68) | 0.48 | 0.48 | 0.44 |
| $\mu = (0, 1), \gamma = (0.4, 0.5)$ | 0.26, (0.76, 0.63) | 0.35, (0.73, 0.66) | 0.38, (0.72, 0.67) | 0.53 | 0.48 | 0.43 |
| $\mu = (0, 2), \gamma = (0.4, 0.5)$ | 0.85, (0.88, 0.81) | 0.88, (0.88, 0.82) | 0.94, (0.87, 0.83) | 0.78 | 0.72 | 0.22 |
| $\mu = (0, 3), \gamma = (0.4, 0.5)$ | 1.47, (0.96, 0.92) | 1.48, (0.96, 0.92) | 1.55, (0.95, 0.93) | 0.91 | 0.88 | 0.09 |

The skewed normal biomarker results are shown in Figures S5–S8. Results are shown for the skewed normally distributed data with no correlation, $\rho = 0$, that compare Marker 1 with $(\mu, \gamma) = (0.55, 0.4)$ to Marker 2 with any specified mean (these results are from Marker 2 with $(\mu, \gamma) = (1, 0.5)$, but the results are the same for any specified mean for Marker 2), Marker 2 with $(\mu, \gamma) = (1, 0.5)$ compared to Marker 1 with $(\mu, \gamma) = (0.55, 0.4)$, Marker 2 with $(\mu, \gamma) = (2, 0.5)$ compared to Marker 1 with $(\mu, \gamma) = (0.55, 0.4)$, and Marker 2 with $(\mu, \gamma) = (3, 0.5)$ compared to Marker 1 with $(\mu, \gamma) = (0.55, 0.4)$. Figure S5 shows the RB of the methods. When there is less separation between disease and control groups such as Marker 1 and the smaller mean of Marker 2, both Youden approaches have the largest RB. As the sample size increases the RB decreases, and when the sample size is imbalanced the RB is about the same as when the sample size is the smallest. The ER-GAM and Concordance-GAM have the smallest RB, with ER similar and little larger than its GAM counterpart and Concordance has larger RB when $\mu \geq 2$ and $n \leq 100$. As the mean for Marker 2 increases ($\mu \geq 2$) Youden-GAM has smaller RB and is the same as the other GAM methods. ER-GAM tends to have the smallest RB.

Figure S6 shows the MSE for the skewed normal markers. The MSE is the largest for Youden. As the sample size increases the MSE decreases, and when imbalanced the MSE is about the same as when the sample size is the smallest. When there is less separation between the disease and control group, such as Marker 1 and the smaller mean of Marker 2, both Youden approaches have the largest MSE. The ER-GAM and Concordance-GAM have the smallest MSE, ER-GAM being the smallest. ER also performs similarly to the GAM approaches. When $\mu \geq 2$ the Youden-GAM performs better and has similar MSE to the Concordance-GAM.

Figure S7 shows the 95% coverages for the methods. The coverages are close to 95%, except when the mean is large the coverage drops a little for Concordance when the sample size is small or imbalanced and ER-GAM with *n* = 200. Figure S8 shows the frequency of cut-offs across methods. Findings are similar to when the data are normally distributed except the Youden-GAM mostly finds a unique cut-off with a small percentage of the time finding two cut-offs. Youden has the largest number of cut-offs. As the sample size increases more cut-offs are identified, particularly with the ER-GAM, Concordance-GAM, and Youden-GAM approaches. As the mean of the markers Increases the number of cut-offs reduces for the ER-GAM, Concordance-GAM, and Youden-GAM approaches but slightly increases for the Concordance and ER approaches.

The skewed normally distributed markers showed the ER-GAM and Concordance-GAM methods to have the best performance overall. The ER-GAM has the best performance, the ER being the next best choice if there is concern about frequency of cut-offs. The Youden-GAM performs the best when the mean is larger for the second biomarker. Regarding the empirical approaches, the ER has the best performance and either the ER or ER-GAM are recommended. A transformation can be done if one is sufficient and then the methods can be tried for normally distributed approaches. We wanted to demonstrate what to use if a transformation is not found or one would rather leave the data in its original scale. As with the normally distributed data, the correlations did not have much impact on the results for the empirical and non-empirical approaches for the skewed distributions.

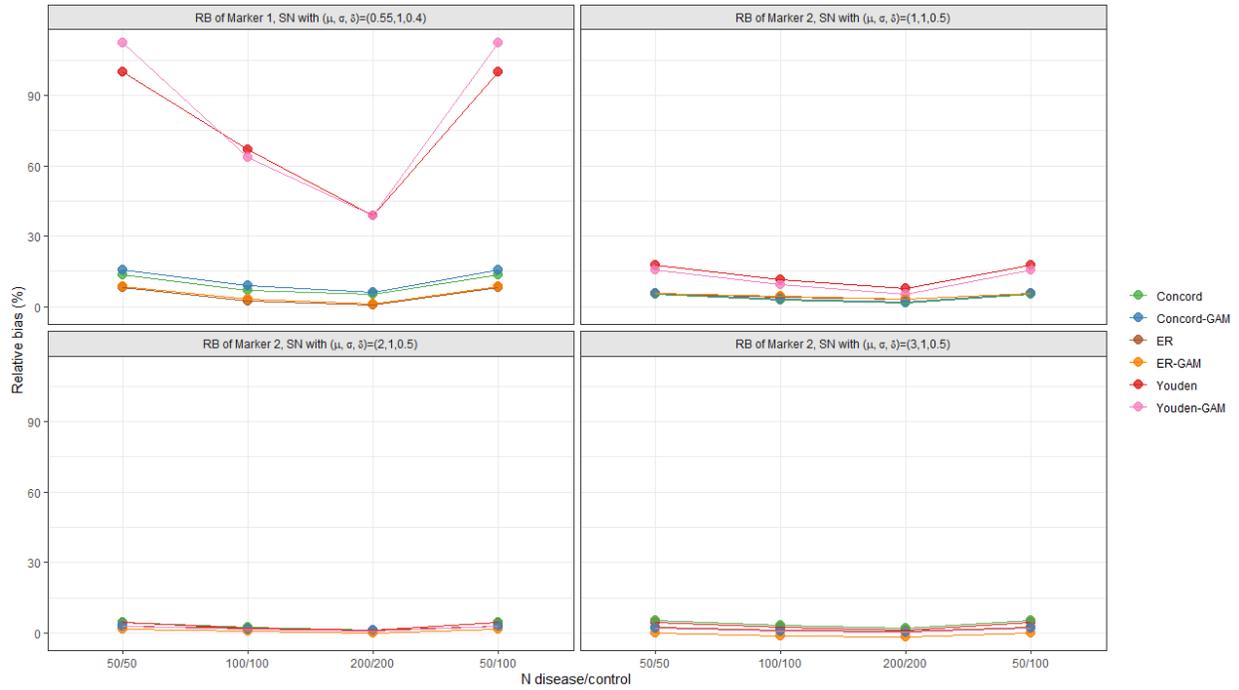

**Figure S5.** Relative bias of skewed normal distributed markers with no correlation between Marker 1 and 2.

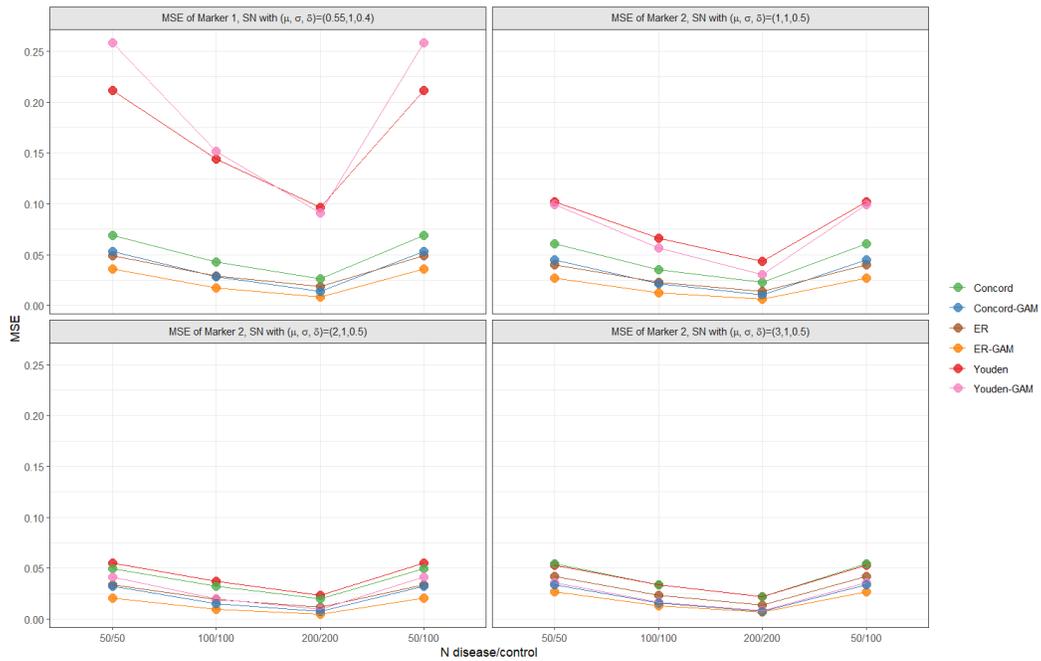

**Figure S6.** MSE of skewed normal distributed markers with no correlation between Marker 1 and 2.

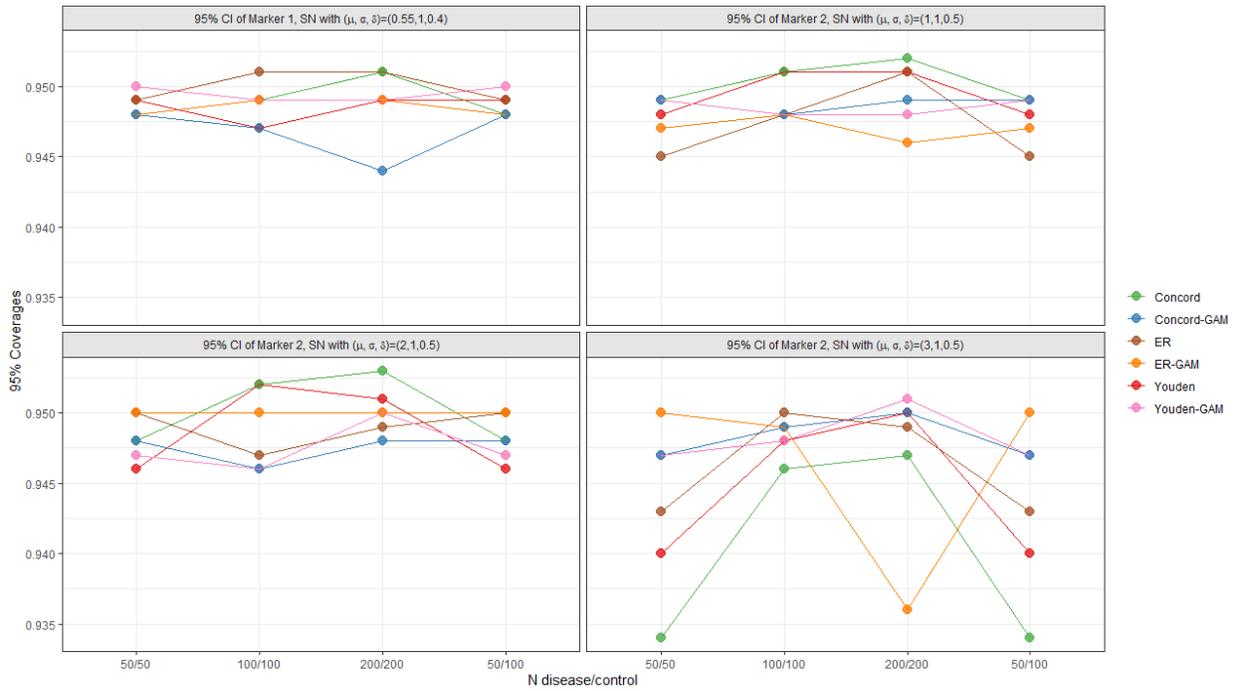

**Figure S7.** Coverages of skewed normal distributed markers with no correlation between Marker 1 and 2.

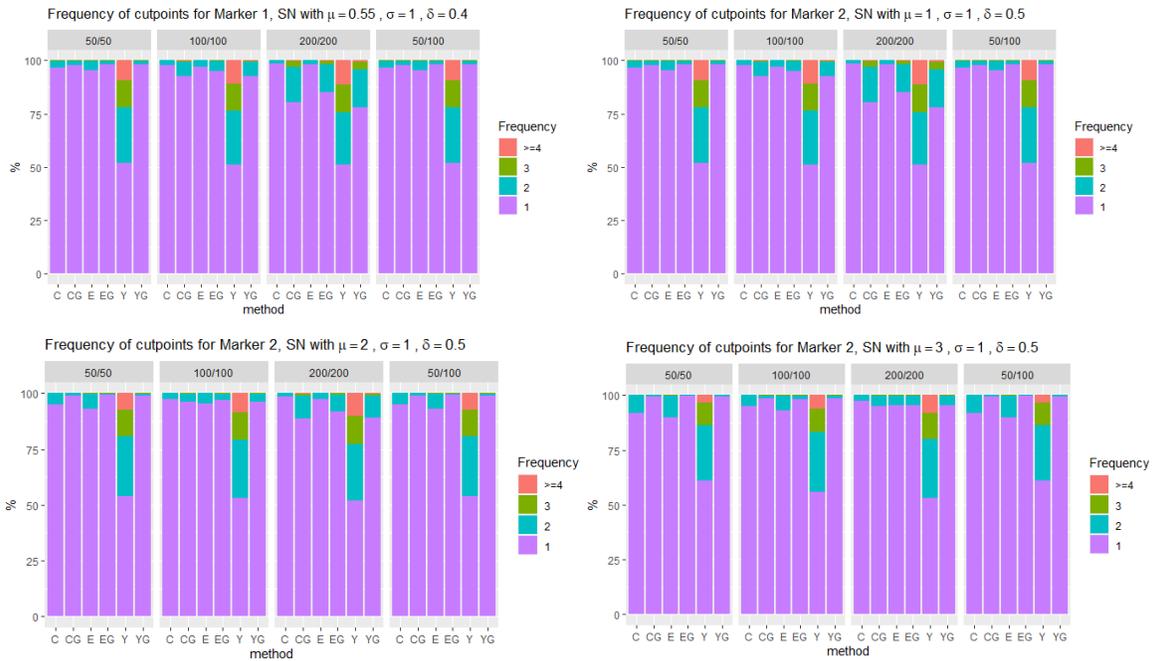

**Figure S8.** Frequency of cut-offs when both markers have skewed normal distribution with no correlation. Methods: C= Concordance, CG= Concordance-GAM, E= ER, EG= ER-GAM, Y=Youden, YG=Youden-GAM.

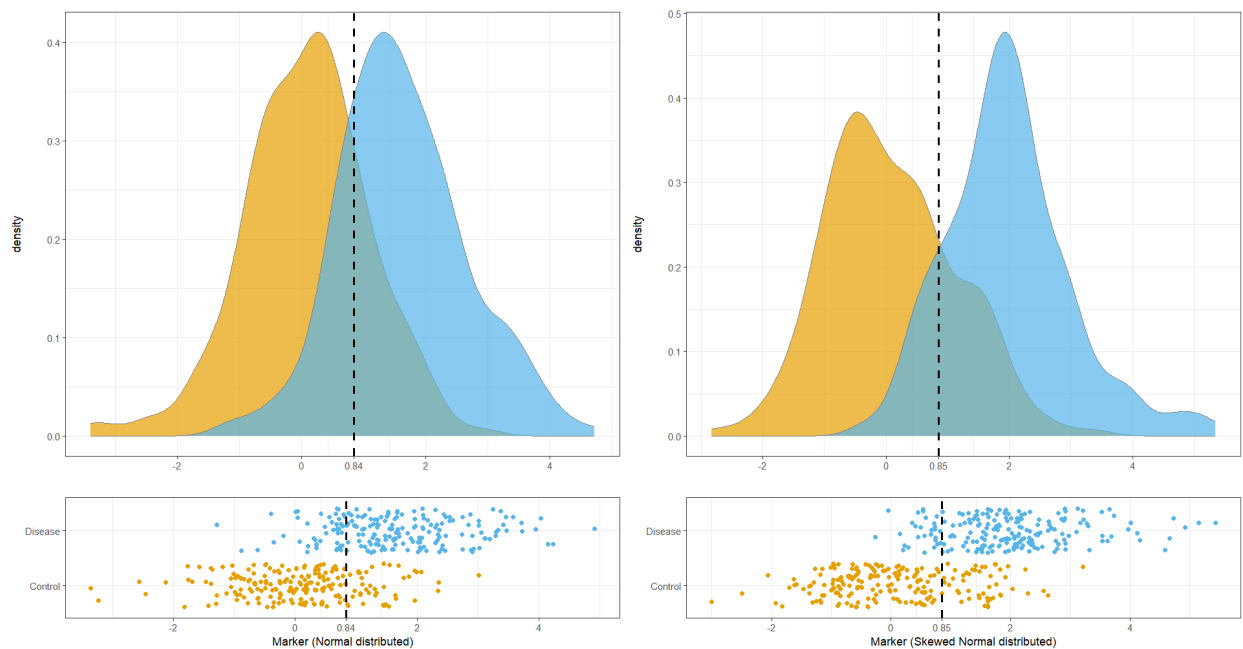

**Figure S9.** Histogram of biomarkers by disease/control group for normally distributed data and skewed normal data with cut-off.

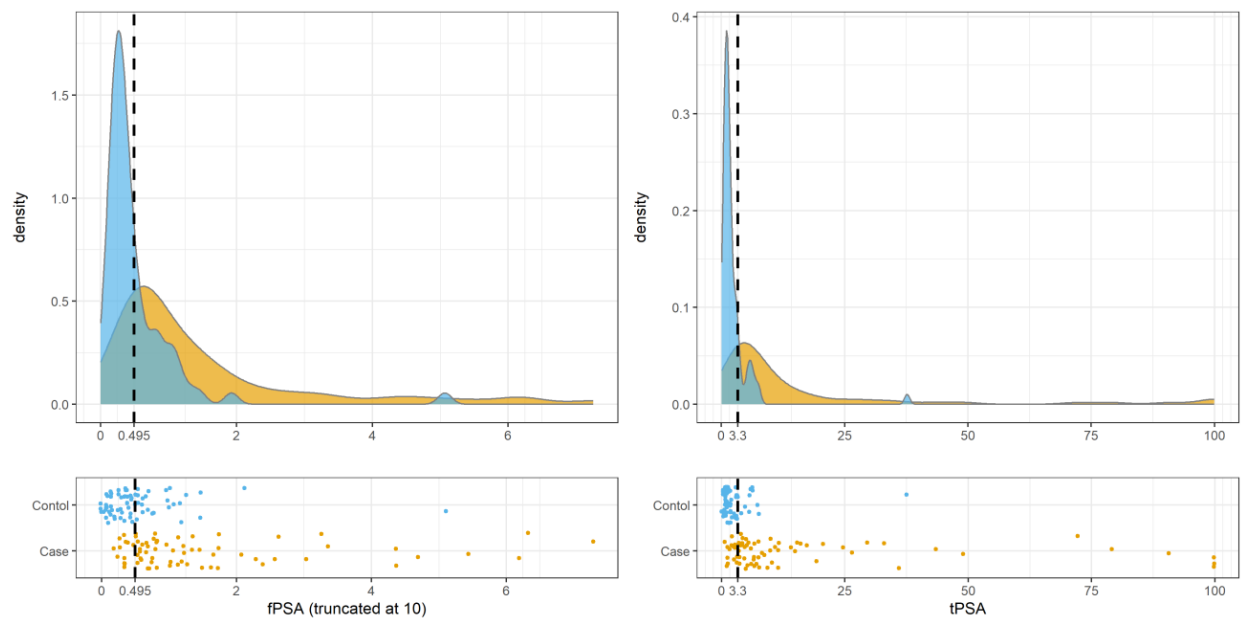

**Figure S10.** Histogram of biomarkers by disease and control groups showing cut-offs in the prostate cancer study example.

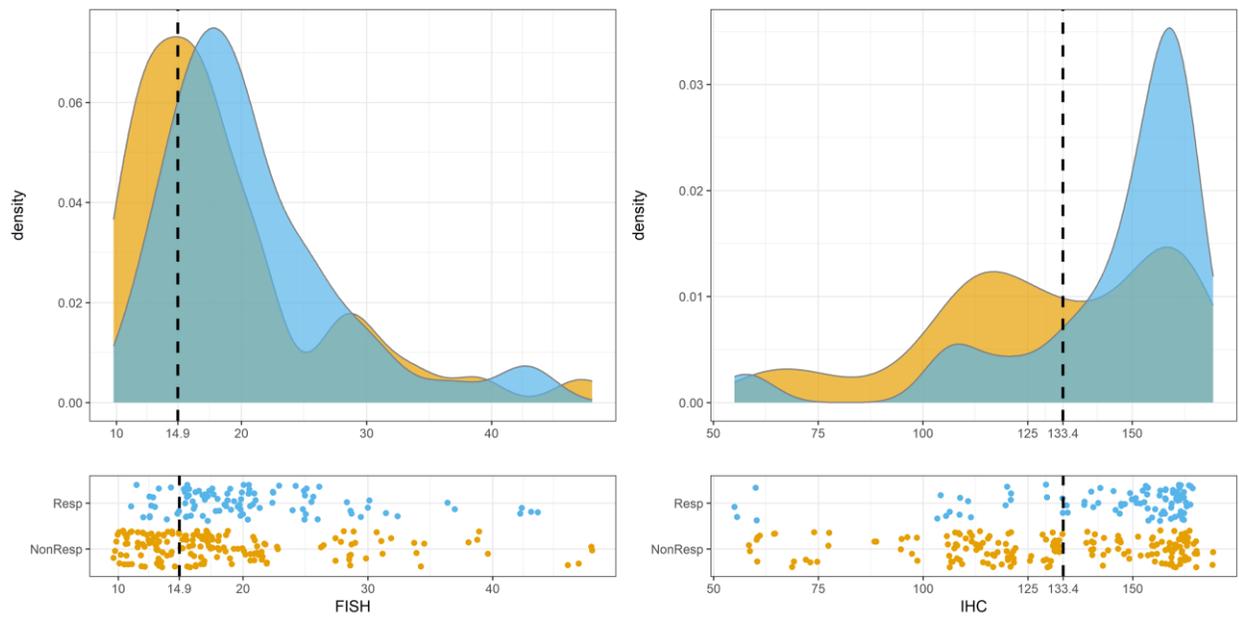

**Figure S11.** Histogram of biomarkers by responder and nonresponder groups showing cut-offs in the lung cancer study example.